\documentclass[twoside]{ae100prg}
\usepackage{graphicx}
\usepackage[breaklinks]{hyperref}
\usepackage{booktabs}
\usepackage{amsmath}

\begin{document}
\renewcommand{\footskip}{10pt}

\title{Energy Extraction from Spinning Black Holes via Relativistic Jets}


\author{Ramesh Narayan$^1$, Jeffrey E.~McClintock$^1$ and 
Alexander Tchekhovskoy$^2$}

\address{$^1$ Harvard-Smithsonian Center for Astrophysics, Harvard
  University, 60 Garden St, Cambridge, MA 02138, USA} 
\address{$^2$ Jadwin Hall,
  Princeton University, Princeton, NJ 08544, USA; Center for
  Theoretical Science Fellow}

\email{rnarayan@cfa.harvard.edu, jmcclintock@cfa.harvard.edu,
atchekho@princeton.edu}

\begin{abstract}

It has for long been an article of faith among astrophysicists that
black hole spin energy is responsible for powering the relativistic
jets seen in accreting black holes. Two recent advances have
strengthened the case.  First, numerical general relativistic
magnetohydrodynamic simulations of accreting spinning black holes show
that relativistic jets form spontaneously.  In at least some cases,
there is unambiguous evidence that much of the jet energy comes from
the black hole, not the disk. Second, spin parameters of a number of
accreting stellar-mass black holes have been measured. For ballistic
jets from these systems, it is found that the radio luminosity of
the jet correlates with the spin of the black hole.  This
suggests a causal relationship between black hole spin and jet power,
presumably due to a generalized Penrose process.

\end{abstract}

\section{Introduction}

Relativistic jets are a common feature of accreting black holes (BHs).
They are found in both stellar-mass BHs and supermassive BHs, and are
often very powerful. Understanding how jets form and where they obtain
their enormous power is an active area of research in astrophysics.

In seminal work, Penrose \cite{Penrose1969} showed that a spinning BH
has free energy that is, in principle, available to be tapped.  This
led to the popular idea that the energy source behind relativistic
jets might be the rotational energy of the accreting BH. A number of
astrophysical scenarios have been described in which magnetic fields
enable this process
\cite{rw75,bz77,damouretal78,koideetal02,komissarov09,
  mckinneygammie04, mckinney06, beskin2010, tchekhovskoyetal11,
  meier12}. Field lines are kept confined around the BH by an
accretion disk, and the rotation of space-time near the BH twists
these lines into helical magnetic springs which expand under their own
pressure, accelerating any attached plasma. Energy is thereby
extracted from the spinning BH and is transported out along the
magnetic field, making a relativistic jet. Although this mechanism
requires accretion of magnetized fluid and is thus not the same as
Penrose's original proposal\footnote{Penrose considered a simple model
  in which particles on negative energy orbits fall into a spinning
  BH. \cite{wd89} extended the analysis to discrete particle accretion
  in the presence of a magnetic field, which introduces additional
  interesting effects.  We do not discuss these particle-based
  mechanisms, but focus purely on fluid dynamical processes within the
  magnetohydrodynamic (MHD) approximation. We also do not discuss an
  ongoing controversy on whether or not different mechanisms based on
  magnetized fluids differ from one another \cite{komissarov09}.}, we
will still refer to it as the ``generalized Penrose process'' since
ultimately the energy comes from the spin of the BH.

It is not easy to prove that the generalized Penrose process is
necessarily in operation in a given jet. The reason is that jets are
always associated with accretion disks, and the accretion process
itself releases gravitational energy, some of which might flow into
the jet.  Let us define a jet efficiency factor $\eta_{\rm jet}$,
\begin{equation}
\eta_{\rm jet} = \frac{\langle L_{\rm jet}\rangle}{\langle\dot{M} (r_{\rm
  H})\rangle c^2},
\end{equation}
where $\langle L_{\rm jet}\rangle$ is the time-average 
(kinetic and electromagnetic) luminosity flowing
out through the jet and $\langle\dot{M}(r_{\rm H})\rangle c^2$ is the
time-average rate at which rest-mass energy flows in through the BH
horizon.  Many jets, both those observed and those seen in computer
simulations, have values of $\eta_{\rm jet}$ quite a bit less than
unity.  With such a modest efficiency, the jet power could easily come
from the accretion disk \cite{bp82, ga97, lop99}.

The situation has improved considerably in the last couple of years.
As we show in \S\ref{sec:simulation}, numerical simulations have now
been carried out where it can be demonstrated beyond reasonable doubt
that the simulated jet obtains power directly from the BH spin energy.
Furthermore, as we discuss in \S\ref{sec:empirical}, the first
observational evidence for a possible correlation between jet power and BH spin
has finally been obtained. The correlation appears to favor a
Penrose-like process being the energy source of jets.

\section{Computer Simulations of Black Hole Accretion and Jets}
\label{sec:simulation}

For the last decade or so, it has been possible to simulate
numerically the dynamics of MHD accretion flows in the fixed Kerr
metric of a spinning BH. The dynamics of the magnetized fluid are
described using the general relativistic MHD (GRMHD) equations in a
fixed space-time, and the simulations are carried out in 3D in order
to capture the magnetorotational instability (MRI), the agency that
drives accretion \cite{balbus_hawley98}. Radiation is usually ignored,
but this is not considered a problem since jets are usually found in
systems with geometrically thick accretion disks, which are
radiatively inefficient. We describe here one set of numerical
experiments \cite{tchekhovskoyetal11, tchekhovskoyetal12} which have
been run using the GRMHD code HARM \cite{gammieetal03} and which are
particularly relevant for understanding the connection between the
generalized Penrose process and jets.

\begin{figure}
  \centering
  \includegraphics[width=\columnwidth]{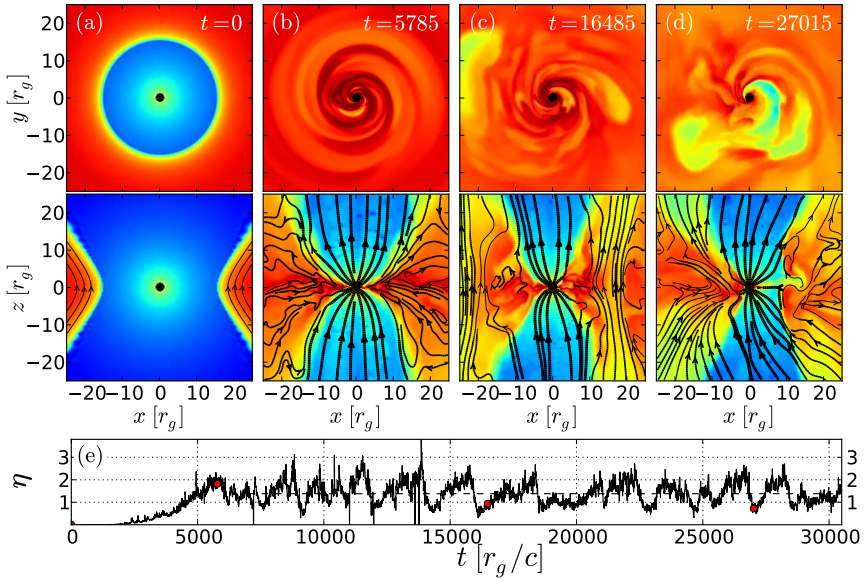}
  \caption{Formation of a magnetically-arrested disk and ejection of
    powerful jets in a GRMHD simulation of magnetized accretion on to
    a rapidly spinning BH with $a_*=0.99$ \cite{tchekhovskoyetal11}.
    The top and bottom rows in panels (a)-(d) show a time sequence of
    equatorial and meridional slices through the accretion flow.
    Solid lines show magnetic field lines in the image plane, and
    color shows $\log\rho$ (red high, blue low).  The simulation
    starts with an equilibrium torus embedded with a weak magnetic
    field (panel a).  The weakly magnetized orbiting gas is unstable
    to the MRI, which causes gas and field to accrete.  As large-scale
    magnetic flux accumulates at the center, a coherent bundle of
    field lines forms at the center, which threads the BH and has the
    configuration of bipolar funnels along the (vertical) BH rotation
    axis. These funnels contain strong field and low mass density
    (lower panels b, c, d). Helical twisting of the field lines as a
    result of dragging of frames causes a powerful outflow of energy
    through the funnels in the form of twin jets. The outflow
    efficiency $\eta$ (panel e), calculated as in
    equation~\eqref{eq:eta}, becomes greater than unity once the flow
    achieves quasi-steady state at time $t\gtrsim 5000r_g/c$.  This is
    the key result of the simulation. Having a time-average $\eta>1$
    means that there is a net energy flow \emph{out} of the BH, i.e.,
    spin energy is extracted from the BH by the magnetized accretion
    flow. This constitutes a demonstration of the generalized Penrose
    process in the astrophysically relevant context of a magnetized
    accretion flow.}
\label{fig:mad}
\end{figure}

As is standard, the numerical simulations are initialized with an
equilibrium gas torus orbiting in the equatorial plane of a spinning
BH. The torus is initially embedded with a weak seed magnetic field,
as shown in panel (a) of Figure~\ref{fig:mad}. Once the simulation
begins, the magnetic field strength grows as a result of the MRI
\cite{balbus_hawley98}. This leads to MHD turbulence, which in turn
drives accretion of mass and magnetic field into the BH. The mass
accretion rate is given by
\begin{equation}
  \label{eq:mdot}
  \dot M(r) = -\iint_{\theta,\varphi}\rho u^r dA_{\theta,\varphi},
\end{equation}
where the integration is over all angles on a sphere of radius $r$
(Boyer-Lindquist or Kerr-Schild coordinates), $dA_{\theta,\varphi} =
\sqrt{-g}d\theta d\varphi$ is the surface area element, $\rho$ is the
density, $u^r$ is the contravariant radial component of 4-velocity,
and $g$ is the determinant of the metric. The sign in
equation~\eqref{eq:mdot} is chosen such that $\dot M > 0$ corresponds
to mass inflow.  One is usually interested in the mass accretion rate
at the horizon, $\dot{M}(r=r_{\rm H})$.  In computing $\dot{M}(r_{\rm
  H})$ from simulations, one waits until the system has reached
approximate steady state. One then computes $\dot{M}(r_{\rm H})$ over
a sequence of many snapshots in time and then averages to eliminate
turbulent fluctuations. This gives the time-average mass accretion
rate $\langle\dot{M}(r_{\rm H})\rangle$.

Panels (b)-(d) in Figure~\ref{fig:mad} show the time evolution of the
accretion flow and jet in a simulation with BH spin $a_*\equiv
a/M=0.99$, where $M$ is the BH mass \cite{tchekhovskoyetal11}.  The steady accretion of
magnetized fluid causes magnetic field to accumulate in the inner
regions near the BH. After a while, the field becomes so strong that
it compresses the inner part of the otherwise geometrically thick
accretion flow into a thin sheet (panel b). The effect is to obstruct
the accretion flow (panels c and d), leading to what is known as a
magnetically-arrested disk \cite{nia03, igumenshchev08} or a
magnetically choked accretion flow \cite{mtb12}. The strong field
extracts BH spin energy and forms a powerful outflow.  To understand
the energetics, consider the rate of flow of energy,
\begin{equation}
  \label{eq:edot}
  \dot E(r) = \iint_{\theta,\varphi}T^r_t dA_{\theta,\varphi},
\end{equation}
where the stress-energy tensor of the magnetized fluid is
\begin{equation}
T^{\mu}_{\nu} = \left(\rho+u_g+p_g+\frac{b^2}{4\pi}\right)u^\mu u_\nu +
\left(p+\frac{b^2}{8\pi}\right)\delta^\mu_\nu - \frac{b^\mu b_\nu}{4\pi},
\end{equation}
$u_g$ and $p_g$ are the internal energy and pressure of the gas,
$b^\mu$ is the fluid-frame magnetic field $4$-vector (see
\cite{gammieetal03} for the definition), and $b^2 = b^\mu b_\mu$ is
the square of the fluid-frame magnetic field strength. The sign of
equation~\eqref{eq:edot} is chosen such that $\dot E(r)>0$ corresponds
to energy inflow.  Note that $T^r_t$ includes the inflow of rest mass
energy via the term $\rho u^ru_t$.

Let us define the efficiency with which the accreting BH produces
outflowing energy as
\begin{equation}
  \label{eq:eta}
  \eta = \frac{\dot M(r_{\rm H}) c^2-\dot E(r_{\rm H})}{\langle\dot
    M(r_{\rm H})\rangle c^2},
\end{equation}
where we have made $\eta$ dimensionless by normalizing the right-hand
side by the time-average mass energy accretion rate. To understand the
meaning of equation~\eqref{eq:eta}, consider the simple example of gas
falling in radially from infinity, with no radiative or other energy
losses along the way. In this case, we have $\dot{E}(r_{\rm H})
=\dot{M}(r_{\rm H})c^2$, i.e., the gas carries an energy equal to its
rest mass energy into the BH. Hence $\eta=0$, as appropriate for this
example. For a more realistic accretion flow, some energy is lost by
the gas via radiation, winds and jets, and one generally expects the
energy flowing into the BH to be less than the rest mass energy:
$\dot{E}(r_{\rm H}) < \dot{M}(r_{\rm H})c^2$. This will result in an
efficiency $\eta>0$, where $\eta$ measures the ratio of the energy
returned to infinity, $\dot{M}(r_{\rm H})c^2-\dot{E}(r_{\rm H})$, to
the energetic price paid in the form of rest mass energy flowing into
the BH, $\dot{M}(r_{\rm H})c^2$.

Usually, $\dot{E}(r_{\rm H})$ is positive, i.e., there is a net flow
of energy into the BH through the horizon, and $\eta<1$. However,
there is no theorem that requires this.  Penrose's \cite{Penrose1969}
great insight was to realize that it is possible to have
$\dot{E}(r_{\rm H})<0$ (net \emph{outward} energy flow as measured at
the horizon), and thus $\eta>1$.  In the context of an accretion flow,
$\dot{E}(r_{\rm H})<0$ means that, even though rest mass flows
steadily into the BH, there is a net energy flow out of the BH.  As a
result, the gravitational mass of the BH decreases with time. It is
the energy associated with this decreasing mass that enables $\eta$ to
exceed unity. 
Of course, as the BH loses gravitational
mass, it also loses angular momentum and spins down. This can be
verified by considering the angular momentum flux at the horizon,
$\dot{J}(r_{\rm H})$, which may be computed as in
equation~\eqref{eq:edot} but with $T^r_t$ replaced by $T^r_\phi$
(e.g., \cite{pennaetal10}).

Returning to the simulation under consideration,
Figure~\ref{fig:mad}(e) shows the outflow efficiency $\eta$ as a
function of time. It is seen that the average efficiency exceeds unity
once the flow achieves steady state at time $t\gtrsim5000r_g/c$, where
$r_g=GM/c^2$. The outflow thus carries away more energy than the
entire rest mass energy brought in by the accretion flow. This is an
unambiguous demonstration of the generalized Penrose process in the
astrophysically plausible setting of a magnetized accretion flow on to
a spinning BH. Of course, it is not obvious that the energy
necessarily flows out in a collimated relativistic jet, since the
quantity $\eta$ is defined via global integrals (eqs.~\ref{eq:mdot},
\ref{eq:eta}) which do not specify exactly where the outflowing energy
ends up. A more detailed analysis reveals that the bulk of the energy
does indeed go into a relativistic jet, while about $10\%$ goes into a
quasi-relativistic wind \cite{tchekhovskoyetal12}.

\begin{figure}
\centering
\includegraphics[width=2.35in,clip]{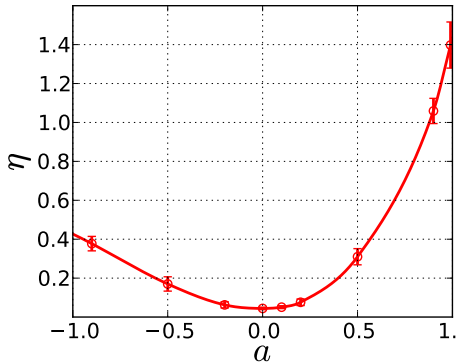}
\caption{Time-average outflow efficiency $\eta$ versus BH spin
  parameter $a_*$ for a sequence of GRMHD simulations of non-radiative
  BH accretion flows \cite{tchekhovskoyetal12}. The efficiency exceeds
  unity for $a_*\gtrsim0.9$. Negative values of $a_*$ correspond to
  the accretion flow counter-rotating with respect to the BH.}
\label{fig:eta}
\end{figure}

Figure \ref{fig:mad} corresponds to an extreme example, viz., a very
rapidly spinning BH with $a_*=0.99$. Figure~\ref{fig:eta} shows
results from a parameter study that investigated the effect of varying
$a_*$. It is seen that the time-average $\eta$ increases steeply with
increasing $a_*$.  For an accretion flow that corotates with the BH,
the power going into the jet can be well-fit with a power-law
dependence,
\begin{equation}
  \label{eq:etafit}
  \eta_{\rm jet} \approx 0.65 a_*^2(1+0.85 a_*^2).
\end{equation}
This approximation remains accurate to within $15\%$ for $0.3\le a_*
\le 1$.  For low spins, the net efficiency derived from the
simulations is greater than that predicted by
equation~\eqref{eq:etafit}. For example, as Figure \ref{fig:eta}
shows, the simulation gives a non-zero value of $\eta$ for $a_*=0$,
which is inconsistent with equation~\eqref{eq:etafit}.  This is
because, for $a_*=0$, all the outflow energy comes directly from the
accretion flow, most of which goes into a wind. Nothing comes from the
BH, whereas equation~\eqref{eq:etafit} refers specifically to the
efficiency $\eta_{\rm jet}$ associated with jet power from the BH.
With increasing BH spin, both the disk and the hole contribute to
energy outflow, with the latter becoming more and more dominant.  For
spin values $a_*>0.9$, the BH's contribution is so large that the net
efficiency exceeds unity. 

Before leaving this topic, we note that
other numerical simulations have used geometrically thicker accretion
configurations than the one shown in Figure \ref{fig:mad} and find
even larger values of $\eta$ \cite{tm12,mtb12}.

\section{Empirical Evidence for the Generalized Penrose Process}
\label{sec:empirical}

As discussed in \S\ref{sec:simulation}, there is definite evidence
from computer simulations that the generalized Penrose process is
feasible, and even quite plausible, with magnetized accretion
flows. We discuss here recent progress on the observational front. In
\S\ref{sec:BHspin} we briefly summarize efforts to measure spin
parameters of astrophysical BHs. Then in \S\ref{sec:jetspin} we
discuss a correlation that has been found between jet power and BH
spin. Finally in \S\ref{sec:Penrose} we explain why we think the
observational evidence favors a Penrose-like process rather than disk
power.

\subsection{Spin Parameters of Stellar-Mass Black Holes}
\label{sec:BHspin}

In 1989, the first practical approach to measuring black hole spin was
suggested \cite{fabianetal89}, viz., modeling the
relativistically-broadened Fe K emission line emitted from the inner
regions of an accretion disk.  The first compelling observation of
such a line was reported six years later \cite{tanakaetal95}.
Presently, the spins of more than a dozen black holes have been
estimated by modeling the Fe K line (see \cite{reynolds13} for a
recent review).

\begin{figure}
\begin{center}
  \includegraphics[width=3in]{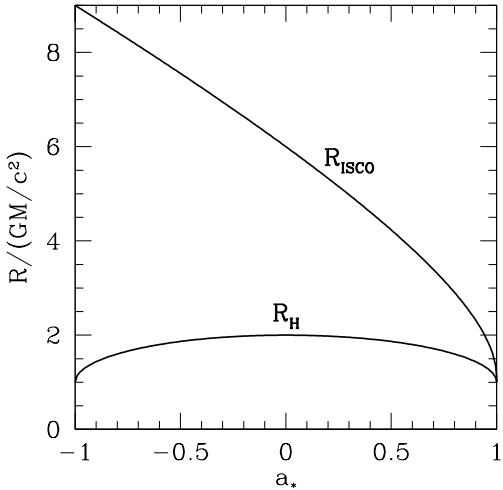}
\end{center}
\vskip -0.7cm
\caption{Radius of the ISCO $R_{\rm ISCO}$ and of the horizon
  $R_{\rm H}$ in units of $GM/c^2$ plotted as a function of the black
  hole spin parameter $a_*$. Negative values of $a_*$ correspond to
  retrograde orbits. Note that $R_{\rm ISCO}$ decreases monotonically
  from $9GM/c^2$ for a retrograde orbit around a maximally spinning
  black hole, to $6GM/c^2$ for a non-spinning black hole, to $GM/c^2$
  for a prograde orbit around a maximally spinning black hole.}
\label{fig:RISCO}
\end{figure}

In 1997, a second approach to measuring black hole spin, the
``continuum-fitting method,'' was proposed \cite{zhangetal97}.  In
this method, one fits the thermal continuum spectrum of a black hole's
accretion disk to the relativistic model of Novikov \& Thorne
\cite{nt73}. One then identifies the inner edge of the modeled disk
with the radius $R_{\rm ISCO}$ of the innermost stable circular orbit
(ISCO) in the space-time metric.  Since $R_{\rm ISCO}$ varies
monotonically with respect to the dimensionless BH spin parameter
$a_*$ (see Fig.~\ref{fig:RISCO}), a measurement of the former
immediately provides an estimate of the latter.

\begin{table}
\caption{The spins and masses of ten stellar-mass black holes
\cite{mcclintocknarayan13}.
\label{tab:results}}
\begin{center}
\footnotesize
\begin{tabular}{lcc}
\hline
\noalign
{\vspace{1mm}}
System & $a_*$ & $M/M_{\odot}$ \\
\noalign
{\vspace{-2.5mm}}
&& \\
\hline
\noalign
{\vspace{1.2mm}}
Persistent && \\
\noalign
{\vspace{1mm}}
\hline
\noalign
{\vspace{1.2mm}}
Cygnus X-1 & $> 0.95$ & $14.8\pm1.0$ \\
{\vspace{-3.5mm}}
&& \\
\noalign
{\vspace{0.9mm}}
LMC X-1 & $0.92_{-0.07}^{+0.05}$ & $10.9\pm1.4$ \\
\noalign
{\vspace{0.9mm}}
M33 X-7 & $0.84\pm0.05$ & $15.65\pm1.45$ \\
\noalign
{\vspace{1mm}}
\hline
\noalign
{\vspace{1mm}}
Transient && \\
\noalign
{\vspace{1mm}}
\hline
\noalign
{\vspace{1.2mm}}
GRS 1915+105 & $> 0.95$ & $10.1\pm0.6$ \\
{\vspace{-3.5mm}}
&& \\
\noalign
{\vspace{0.9mm}}
4U 1543--47 & $0.80\pm0.10$ & $9.4\pm1.0$ \\
\noalign
{\vspace{0.9mm}}
GRO J1655--40 & $0.70\pm0.10$ & $6.3\pm0.5$ \\
\noalign
{\vspace{0.9mm}}
XTE J1550-564 & $0.34\pm0.24$ & $9.1\pm0.6$ \\
\noalign
{\vspace{0.9mm}}
H1743-322 & $0.2\pm0.3$ & $\sim8$ \\
\noalign
{\vspace{0.9mm}}
LMC X-3 & $< 0.3$ & $7.6\pm1.6$ \\
\noalign
{\vspace{0.9mm}}
A0620-00 & $0.12\pm0.19$ & $6.6\pm0.25$ \\
\noalign
{\vspace{0.7mm}}
\hline
\end{tabular}
\end{center}
\end{table}

In 2006, the continuum-fitting method was employed to estimate the
spins of three stellar-mass BHs \cite{shafeeetal06, mcclintocketal06}.
Seven additional spins have since been measured. Table
\ref{tab:results} lists the masses and spins of these ten BHs. Readers
are referred to a recent review 
\cite{mcclintocknarayan13} for details of the continuum-fitting method
and uncertainties in the derived spin estimates.

The continuum-fitting method is simple and demonstrably robust. It
does not make many assumptions; those few it makes have nearly all
been tested and shown to be valid (see \cite{mcclintocketal11,
  mcclintocknarayan13} for details).  A significant limitation of the
method is that it is only readily applicable to stellar-mass BHs.  For
such BHs, however, we would argue that it is the method of choice.
The Fe K method can be applied to both stellar-mass and supermassive
BHs.  For the latter, it is the only method currently available.

\subsection{Correlation Between Black Hole Spin and Jet Radio Power}
\label{sec:jetspin}

The 10 stellar-mass BHs in Table \ref{tab:results} are divided into
two classes: ``persistent'' sources, which are perennially bright in
X-rays at a relatively constant level, and ``transient'' sources,
which have extremely large amplitude outbursts. During outburst, the
transient sources generally reach close to the Eddington luminosity
limit (see \cite{steineretal13} for a quantitative discussion of this
point). Close to the peak, these systems eject blobs of plasma that
move ballistically outward at relativistic speeds (Lorentz factor
$\Gamma>2$).  These ballistic jets are often visible in radio and
sometimes in X-rays out to distances of order a parsec from the BH,
i.e., to distances $>10^{10}GM/c^2$.  Because ballistic jets resemble
the kiloparsec-scale jets seen in quasars, stellar-mass BHs that
produce them are called microquasars \cite{mirabelrodriguez99}.

On general principles, one expects the jet luminosity $L_{\rm jet}$
(the power flowing out in the kinetic and electromagnetic energy) to depend on the BH mass
$M$, its spin $a_*$, and the mass accretion rate $\dot{M}$ (plus
perhaps other qualitative factors such as the topology of the magnetic
field \cite{beckwithetal08, mckinney_blandford09}). If one wishes to investigate the dependence of jet luminosity on
$a_*$, one needs first to eliminate the other two variables.
Ballistic jets from transient stellar-mass BHs are very well-suited
for this purpose.  First, the BH masses are similar to better than a
factor of two (see Table \ref{tab:results}). Second, all these sources
have similar accretion rates, close to the Eddington limit, at the
time they eject their ballistic jets \cite{steineretal13}. This leaves
$a_*$ as the only remaining variable.

\begin{figure}
  \includegraphics[width=0.45\textwidth]{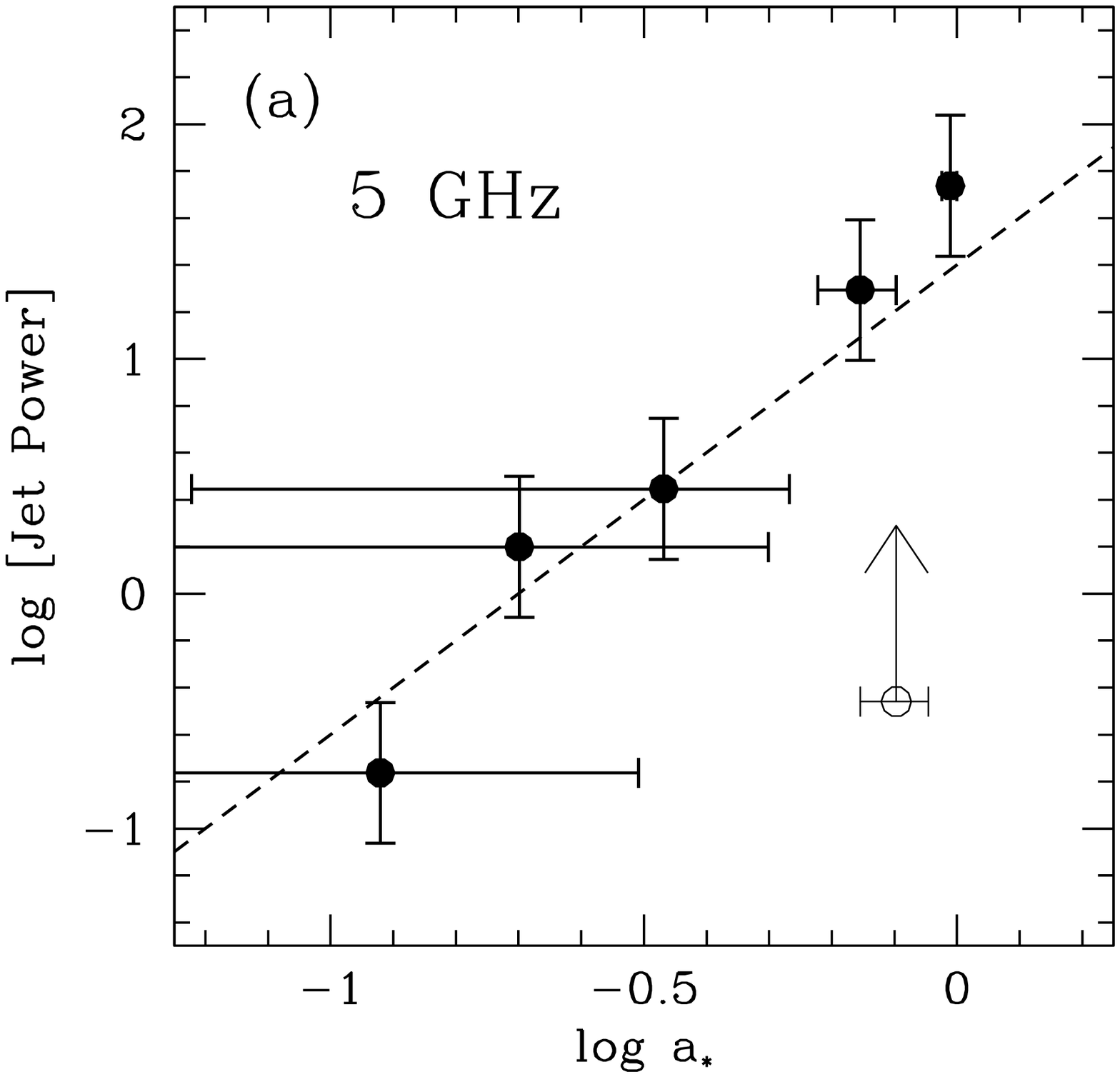}
  \includegraphics[angle=90,width=0.55\textwidth]{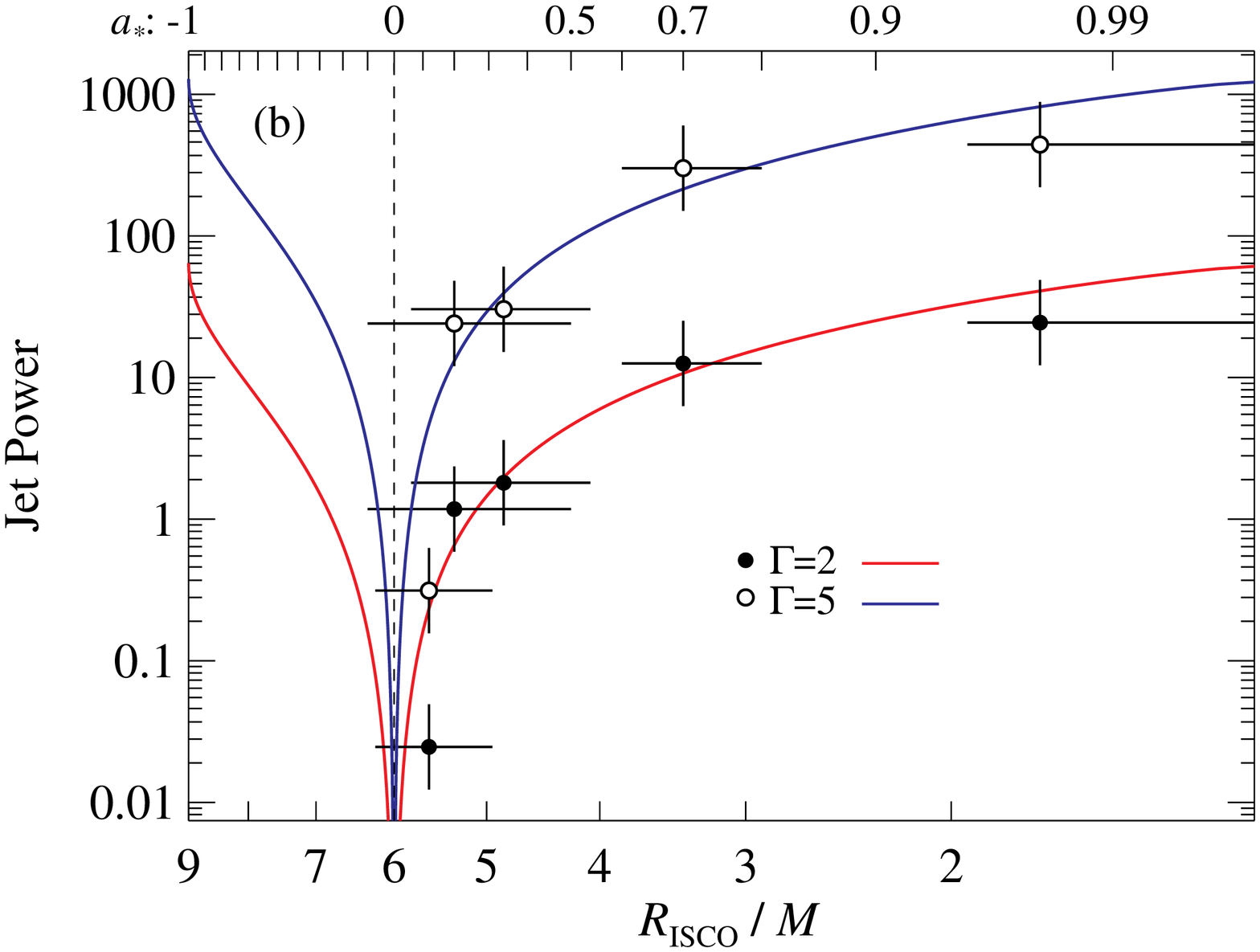}
\caption{(a) Plot of jet power estimated from 5\,GHz radio flux at
  light curve maximum, versus black hole spin measured via the
  continuum-fitting method, for five transient stellar-mass BHs
  \cite{narayanmcclintock12, steineretal13}. The dashed line has slope
  fixed to 2 (see eq.~\ref{eq:etafit}) and is not a fit. (b) Plot of
  jet power versus $R_{\rm ISCO}/(GM/c^2)$.  Here jet power has been
  corrected for beaming assuming jet Lorentz factor $\Gamma=2$ (filled
  circles) or $\Gamma=5$ (open circles). The two solid lines
  correspond to fits of a relation of the form, ``Jet Power'' $\propto
  \Omega_{\rm H}^2$, where $\Omega_{\rm H}$ is the angular frequency
  of the black hole horizon \cite{steineretal13}. Note that jet power
  varies by a factor $\gtrsim1000$ among the five objects shown.}
\label{fig:jet}
\end{figure}

Narayan \& McClintock \cite{narayanmcclintock12} considered the peak
radio luminosities of ballistic jet blobs in four transient BHs,
A0620-00, XTE J1550-564, GRO J1655-40, GRS 1915+105, and showed that
they correlate well with the corresponding black hole spins measured
via the continuum-fitting method\footnote{In the case of a fifth
  transient BH, 4U1543-47, radio observations did not include the peak
  of the light curve, so one could only deduce a lower limit to the
  jet power, which is shown as an open circle in Figure
  \ref{fig:jet}a.}. Later, \cite{steineretal13} included a fifth BH,
H1743-322, whose spin had been just measured. Figure~\ref{fig:jet}a
shows the inferred ballistic jet luminosities of these five objects
plotted versus black hole spin. The quantity ``Jet Power'' along the
vertical axis refers to the scaled maximum radio luminosity $(\nu
S_\nu)D^2/M$, where $\nu$ ($=5$\,GHz) is the radio frequency at which
the measurements are made, $S_\nu$ is the radio flux density in
janskys at this frequency at the peak of the ballistic jet radio light
curve, $D$ is the distance in kiloparsecs, and $M$ is the black hole
mass in solar units. There is unmistakeable evidence for a correlation
between the observationally determined quantity Jet Power and
$a_*$. Although there are only five data points, note that Jet Power
varies by nearly three orders of magnitude as the spin parameter
varies from $\approx0.1-1$.

The very unequal horizontal errorbars in Figure~\ref{fig:jet}a are a
feature of the continuum-fitting method of measuring $a_*$. Recall
that the method in effect measures $R_{\rm ISCO}$ and then deduces the
value of $a_*$ using the mapping shown in Figure~\ref{fig:RISCO}.
Since the mapping is non-linear, especially as $a_*\to1$, comparable
errors in $R_{\rm ISCO}$ correspond to very different uncertainties in
$a_*$. In addition, the use of $\log a_*$ along the horizontal axis
tends to stretch errorbars excessively for low spin values.
Figure~\ref{fig:jet}b, based on \cite{steineretal13}, illustrates
these effects.  Here the horizontal axis tracks $\log R_{\rm ISCO}$
rather than $\log a_*$, and the horizontal errorbars are therefore
more nearly equal. The key point is, regardless of how one plots the
data, the correlation between Jet Power and black hole spin appears to
be strong.

\subsection{Why Generalized Penrose Process?}
\label{sec:Penrose}

Assuming the correlation shown in Figure~\ref{fig:jet} is real, there
are two immediate implications: (i) Ballistic jets in stellar-mass BHs
are highly sensitive to the spins of their underlying BHs.  (ii) Spin
estimates of stellar-mass BHs obtained via the continuum-fitting
method are sufficiently reliable to reveal this long-sought connection
between relativistic jets and BH spin.

With respect to (i), the mere existence of a correlation does not
necessarily imply that the generalized Penrose process is at work. We
know that the accretion disk itself is capable of producing a jet-like
outflow \cite{bp82, ga97, lop99}. Furthermore, the gravitational
potential well into which an accretion disk falls becomes deeper with
increasing BH spin, since the inner radius of the disk $R_{\rm ISCO}$
becomes smaller (Fig.~\ref{fig:RISCO}). Therefore, a disk-driven jet
is likely to become more powerful with increasing spin. Could this be
the reason for the correlation between Jet Power and spin seen in
Figure~\ref{fig:jet}?  We consider it unlikely. The radiative
efficiency $\eta_{\rm disk}$ of a Novikov-Thorne thin accretion disk
increases only modestly with spin; for the spins of the five objects
shown in Figure~\ref{fig:jet}, $\eta_{\rm disk}=$ 0.061, 0.069, 0.072,
0.10 and 0.19, respectively, varying by only a factor of three. Of
course, there is no reason why the power of a disk-driven jet should
necessarily scale like $\eta_{\rm disk}$. Nevertheless, the fact that
$\eta_{\rm disk}$ shows only a factor of three variation makes it
implausible that a disk-powered jet could have its radio luminosity
vary by three orders of magnitude.

In contrast, any mechanism that taps directly into the BH spin energy
via some kind of generalized Penrose process can easily account for
the observed variation in Jet Power. Analytical models of magnetized
accretion predict that the jet efficiency factor should vary as
$\eta_{\rm jet}\propto a_*^2$ \cite{rw75, bz77} or $\eta_{\rm
  jet}\propto \Omega_{\rm H}^2$ \cite{tchekhovskoyetal10}, where
$\Omega_{\rm H}$ is the angular frequency of the BH
horizon\footnote{The two scalings agree for small values of $a_*$, but
  differ as $a_*\to1$.}. The dashed line in Figure~\ref{fig:jet}a
corresponds to the former scaling, and the solid lines in
Figure~\ref{fig:jet}b to the latter scaling;
equation~\eqref{eq:etafit}, which is obtained by fitting simulation
results, is intermediate between the two.  The observational data
agree remarkably well with the predicted scalings, strongly suggesting
that the generalized Penrose process is in operation.

We cannot tell whether the energy extraction in the observed systems
is mediated specifically by magnetic fields (as in the simulations),
since there is no way to observe what is going on near the BH at the place
where the jet is initially launched.  Where the ballistic jet blobs are
finally observed they are clearly magnetized --- it is what enables
the charged particles to produce radiation via the synchrotron
mechanism --- but this is at distances $\sim 10^{10}GM/c^2$.

\section{Summary}

In summary, the case for a generalized version of the Penrose process
being the power source behind astrophysical jets has become
significantly stronger in the last few years. Computer simulations
have been very helpful in this regard since they enable one to study
semi-realistic configurations of magnetized accretion flows and to
explore quantitatively how mass, energy and angular momentum flow
through the system. Recent computer experiments find unambiguous
indications for energy extraction from spinning BHs via magnetic
fields. Whether these simulated models describe real BHs in nature is
not yet certain. However, completely independent observational data
suggest a link between the spins of transient stellar-mass BHs and the
energy output in ballistic jets ejected from these systems. The
observationally measured quantity ``Jet Power'' in
Figure~\ref{fig:jet} increases steeply with the measured BH spin, and
the dependence is quite similar to that found both in simple
analytical models \cite{rw75, bz77} and in simulations
(Fig.~\ref{fig:eta}).  Taking all the evidence into account, the
authors believe that Penrose's seminal ideas on energy extraction from
spinning BHs are relevant for the production of at least some
categories of relativistic astrophysical jets.

The authors thank J.~F.~Steiner for help with Fig.~\ref{fig:jet}b.
RN's work was supported in part by NASA grant NNX11AE16G.  AT was
supported by a Princeton Center for Theoretical Science fellowship and
an XSEDE allocation TG-AST100040 on NICS Kraken and Nautilus and TACC
Ranch.

\end{document}